\documentclass{emulateapj}

\begin{document}
\bibliographystyle{apj}
\title[Magnetic Fields Line Diffusion]{Asymmetric Diffusion of Magnetic Field Lines}
\author{Andrey Beresnyak}
\affil{Los Alamos National Laboratory, Los Alamos, NM, 87545}

\begin{abstract}
 Stochasticity of magnetic field lines is important for particle transport properties.
 Magnetic field lines separate faster than diffusively in turbulent plasma, which is called superdiffusion.
 We discovered that this superdiffusion is pronouncedly asymmetric, so that the separation of field lines
 along the magnetic field direction is different from the separation in the opposite direction.
 While the symmetry of the flow is
 broken by the so-called imbalance or cross-helicity, the difference between forward and backward diffusion
 is not directly due to imbalance, but a non-trivial consequence of both imbalance and non-reversibility
 of turbulence.
 The asymmetric diffusion perpendicular to the mean magnetic field entails a variety of new physical phenomena,
 such as the production of parallel particle streaming in the presence of perpendicular particle gradients.
 Such streaming and associated instabilities could be significant for particle transport in laboratory, space
 and astrophysical plasmas.
\end{abstract}

\maketitle

\section{Introduction} 
Astrophysical plasmas feature a huge separation between the energy containing scale and the dissipation scale.
Such a high-Reynolds number flows are necessarily turbulent \citep[see, e.g.,][]{Armstrong1995}. 
Conductive turbulent fluids generate their own magnetic fields by dynamo, as a result of which
most of the astrophysical objects are magnetized to some degree, typically close to equipartition between
kinetic and magnetic energy. One of the consequences of turbulence, whether in laboratory, astrophysical or laboratory plasmas,
is the magnetic field line stochasticity, which plays a crucial role in most key physical processes, such as thermal conduction,
reconnection, particle transport, etc. Self-excited turbulence in tokamaks results in anomalous transport of particles perpendicular to
the field and prevents reaching higher temperatures and densities. Recently, unusual features in the angular
distribution of the arrival directions of cosmic rays \citep{Abbasi2011} and spectral features for positrons \citep{adriani2009}
revived studies of cosmic ray propagation in stochastic magnetic fields \citep[see, e.g.,][]{BYL11,Kistler2012}. 
Indeed, carefully studying cosmic ray diffusion could help discriminate between different models of positron excess,
Fermi bubbles and other currently unexplained cosmic ray phenomena.  

Turbulence is now understood as a multi-scale phenomenon, by large owing to the pioneering paper of \citet{richardson1926},
who studied turbulent diffusion and suggested that the diffusion coefficient depends of scale as $D \sim l^{4/3}$, known as Richardson's law.
Indeed, if two particles are separated by a distance $l$, the typical separation speed corresponds to the typical turbulent velocity on scale $l$, which
is, approximately, $\delta v \sim l^{1/3}$ \citep{kolm41}. This suggests that separation between
particles grow as $\delta l \sim t^{2/3}$. If turbulence is uniform and characterized only by the dissipation rate per unit mass, $\epsilon$,
which has units of ${\rm cm}^2/{\rm s}^3$, it is natural that the separation between two particles moving with the fluid, $\Delta {\bf r}$,
conforms to $(\Delta {\bf r})^2=g_0 \epsilon t^3$, where $g_0$ is a dimensionless number known as Richardson's constant.
Richardson's diffusion has been studied extensively by experimental, theoretical and numerical means \citep[see, e.g.,][and refs therein]{sawford2008}.
The turbulent diffusion of particles embedded in the MHD fluid received relatively less attention, however, the same type of diffusion
is expected in perpendicular direction due to the same scaling $\delta v \sim l^{1/3}$ of strong MHD turbulence \citep[see, e.g.,][]{GS95, B11, B12b}.

A different and very interesting question is how the magnetic field lines
separate from each other in such an environment. This question is crucial because well-magnetized plasmas are often poorly
collisional, with the ion Larmor radius being many orders of magnitude smaller that the mean free path from Coulomb collisions. In particular, the magnetized solar
wind features mean free paths which are comparable to the distance from the Sun. In galaxy clusters the Coulomb mean free path is around 10-100kpc.
In most astrophysical environments there is also a high-energy component, called cosmic rays, for which Coulomb collisions are essentially negligible.
Charged particles will, therefore, move along magnetic field lines for great distances and scatter mostly by magnetic perturbations. This will result in
parallel diffusion being much larger than the perpendicular diffusion. In the absence of perpendicular momentum, particle will move along
magnetic field lines and diffuse only due to the magnetic field line diffusion. The motion of the bulk of the plasma $\delta v$ that causes
ordinary diffusion could be neglected if the ion speed $v_i$ is much larger than $\delta v$ \footnote{This is equivalent to the condition that the motions in the inertial range
are subsonic. While some of the astrophysical turbulence feature supersonic motions on the outer scale, the inertial range motions are normally
subsonic.}. For cosmic rays this condition is also very well satisfied, because they move along field lines with the speed comparable to the speed
of light $c$. In other words, at least for short timescales, the fluid is frozen from most particles' perspective.

Despite being collisionless, plasmas in many circumstance can be described as fluids on scales larger that the ion Larmor radius \citep{schekochihin2009}.
The inertial range of MHD turbulence feature strongly anisotropic perturbations which are much smaller in amplitude that the mean magnetic field.
The key component of this turbulence is Alfvenic mode, which is why it is often called Alfvenic turbulence. Due to the fact that the Alfven mode
is driven by magnetic tension, not pressure, it is relatively unaffected by the lack of collisions. The presence of slow mode
in such highly anisotropic turbulence neither affect dynamics \citep{GS95, B12b}, nor influences magnetic field lines, as the anisotropic slow mode perturbation
is mostly along the mean field. Therefore, the equations for the Alfvenic components, which are conventionally called reduced MHD (RMHD), are
sufficient for studying field lines.

\begin{table}[t]
\caption{Simulation parameters}
\begin{tabular*}{0.99\columnwidth}{@{\extracolsep{\fill}}l c c c c r}
    \hline\hline
Run  & Resolution & $\epsilon^+/\epsilon^-$ & $w^+/w^-$  & $\ell_{\|*}^-/\ell_{\|*}^+$ & $g_m^+/g_m^-$\\
   \hline
B1& $1536^3$                &  1     & 1      & 1 & 1 \\
I1 &  $512\cdot 1024^2$  &  1.19 &  1.16 & 1.07 & 1.03 \\
I3 &  $512\cdot 1024^2$  & 1.41  &  1.37   & 1.15 & 1.15 \\
I5 &  $1024\cdot 1536^2$ & 2.00 &  2.36  & 1.36 & 1.31 \\
I6 &  $1024\cdot 1536^2$ & 4.50 &  6.70   & 1.78 & 1.71 \\
   \hline

\end{tabular*}
  \label{imb_experiments}
\end{table}

Perturbations in a strong mean magnetic field could be decomposed into backward and forward propagating eigenmodes ${\bf w^\pm=v\pm b}/\sqrt{4\pi \rho}$ called
Els\"asser variables. Since perturbation sources are not uniform, MHD turbulence is naturally {\it imbalanced}, i.e. the amplitudes of $w^+$ and $w^-$
are not equal. This is verified by direct observations in the solar wind, where the dominant always propagates away from the Sun \citep[see, e.g.,][]{wicks2011}.
Other astrophysical sources are expected to have strong imbalance, for example stellar winds and jets will emit predominantly outward-propagating component. Similarly,
AGN jets are expected to have Alfv\'en perturbations propagating away from the central engine, e.g., due to the black hole spin \citep{blandford1977}.
The theories of imbalanced Alfvenic turbulence are fairly young and has been verified mostly by comparison with simulations \citep{BL09a,B11}, although
the solar wind measurements also show some promise. So far the model most consistent with the data is \citet{BL08}, which correctly explains
the ratio of anisotropies and the ratio of energies, given a certain ratio of energy fluxes. Imbalanced relativistic force-free MHD turbulence, supposedly existing in such
objects as parsec-scale jets and GRB engines has been simulated recently by \citet{Cho2013} and seem to exhibit properties consistent with the \citet{BL08} model.
Since we expect Alfv\'en mode to survive in low-collisional environments such as jets, pulsar winds, etc., we are particularly interested
in the properties of magnetic field line diffusion of Alfv\'enic turbulence. The study of magnetic field diffusion is also equivalent to the study of charged
particle diffusion in the limit of negligible pitch angle scattering, e.g., due to zero perpendicular momentum.

\section{Richardson-Alfv\'en diffusion}
Assuming very strong mean field ${\bf B_0}$ pointing in the $x$ direction, the equation for the magnetic field line is
\begin{equation}
\frac{d \bf r}{dx}= \frac{\bf b}{B_0},
\end{equation}
where the magnetic perturbation ${\bf b=B-B_0}$ is perpendicular to the mean field (Alfv\'en mode), so that the displacement
vector $r$ will only have perpendicular components. Similarly, particles moving along such a field in one direction will only experience
perpendicular diffusion, as $dx \gg |d{\bf r}|$. Suppose, we follow magnetic field lines started from two points, separated by a small initial distance $r_0$.
As the difference between ${\bf B}$ scales as $\delta B_l \sim l^{1/3}$ we would expect a stochastic separation of the magnetic field lines
in to follow the law
\begin{equation}
\langle(\Delta {\bf r})^2\rangle=g_{m} \epsilon v_A^{-3} |x|^3,
\end{equation}
where $\epsilon$ is the dissipation rate per unit mass, as defined above, and $v_A=B_0/\sqrt{4\pi \rho}$ is the Alfv\'en speed.
This expression can be obtained by replacing $t$ in Richardson's formula with the time variable for the Alfven wave, $x/v_A-t$ with $t=0$.
For the lack of a better term in the literature we will designate this Richardson-Alfv\'en diffusion and call the
dimensionless constant $g_{m}$ Richardson-Alfv\'en constant. Although we call this diffusion by analogy with physical diffusion
in time, this is rather a stochastic separation in space. Nevertheless, the term diffusion seems appropriate due to similarity with physical diffusion
and relevance of this problem to perpendicular diffusion of particles.

If magnetic field lines separate for a distance much larger than the outer scale of turbulence $L$, the $\delta {\bf b}$ becomes truly
random and independent of the separation. In this limit the magnetic field lines experience random walk, i.e. ordinary
diffusion $\langle(\Delta {\bf r})^2\rangle \sim |x|$. This limit is known as field line random walk (FLRW) and has been used to describe perpendicular
diffusion in \citet{jokipii1973}. Note that the random walk must be symmetric with respect to the sign of $x$.

\section{Numerical Results}

\begin{figure}[t]
\figurenum{1}
\includegraphics[width=1.0\columnwidth]{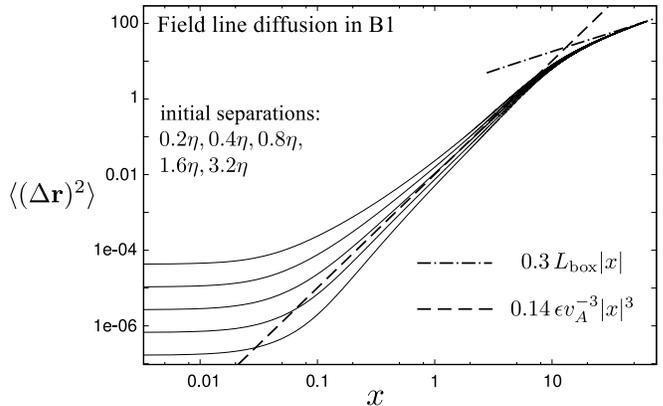}
\caption{Diffusion of magnetic field lines. With initial separations $r_0$ ($0.2\eta \div 3.3\eta$) two field lines start diffusing
apart at $r\sim 12\eta$ by Richardson-Alfv\'en law with $g_{m}=0.14$ and transition to ordinary diffusion with coefficient $0.3 L_{\rm box}$
at separations around outer scale of turbulence.}
\label{rich}
\end{figure}

We used magnetic field snapshots obtained in simulations of Alfv\'enic turbulence. 
These simulations solved the reduced MHD equations with explicit dissipation and driving to achieve
statistically stationary state. Further details behind the RMHD rationale, simulation setup, driving,
numerical scheme, etc., can be found in \citet{B12b}. Each simulation represents stationary,
strong MHD turbulence with strong mean field. The balanced simulation B1 has been previously reported in \citet{B11} and imbalanced
simulations I1-6 has been reported in \citet{BL10}. More details concerning these simulations can be found in the above references.
The parameters of the simulations are summarized in Table~1, with the defining feature of each imbalance simulation being the
ratio of the dissipation rates $\epsilon^\pm$ for Els\"asser components ${\bf w^\pm}$.

We were tracking the pairs of magnetic field lines started at random positions throughout the box and initially separated by distance
$r_0$ by Eq.~1. Fig.~\ref{rich} shows the tracking results
for the B1 simulation. The transition to Richardson's diffusion happens when particles are separated by around twelve Kolmogorov (dissipation) scales $\eta$.
We chose five initial separations, fractions of $\eta$. At large distances they seem to converge towards Richardson-Alfven diffusion with $g_m=0.14$.
At sufficiently large distances the field lines started experiencing random walk, i.e. ordinary diffusion with diffusion coefficient of $0.3 L_{box}$.
We typically used $4\cdot 10^5$ field line pairs for statistical averaging.

\begin{figure}[t]
\figurenum{2}
\includegraphics[width=1.0\columnwidth]{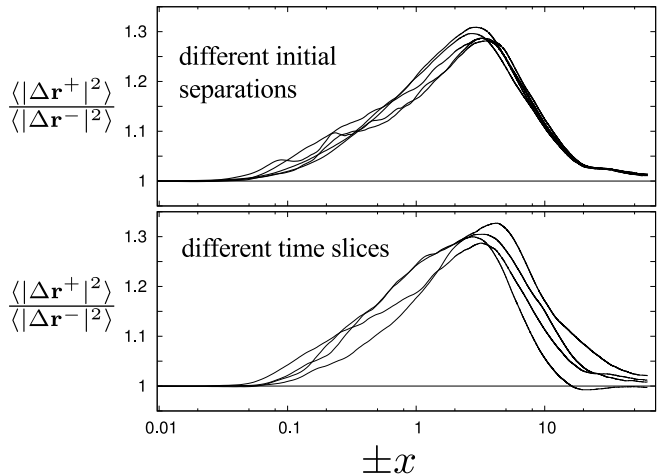}
\caption{The ratio of forward to backward diffusion in datacubes from simulation I5.
Upper plot shows five curves from the same datacube with different initial separations, same as on Fig.~\ref{rich}.
Lower plot shows time variability of the ratio.}
\label{ratio}
\end{figure}

The next tracking experiment involved simulations I1-I6. These also experienced Richardson-Alfv\'en diffusion, but now the diffusion speed was different
depending on whether we track magnetic fields forward or backward (negative or positive $dx$). We plotted the ratio of forward to backward separations
from the simulation I5 on Fig.~\ref{ratio}. It turned out that this ratio is fairly insensitive to the initial separation $r_0$, when $r_0$ was varied by a factor of around 16, with most
difference being due to statistical error. Different snapshots of the same simulation showed more variation. We used this variation in time
to estimate the error in the diffusion ratio. In the large $x$ limit this ratio went to unity, consistent with the symmetry of random walk. 
We estimated the ratio of Richardson-Alfven constants by taking the maximum of the ratio curves, which was also somewhere
around the middle of the inertial range in terms of perpendicular separation. The measurements of the ratio $g_m^+/g_m^-$ are presented in Table~1
and Fig.~\ref{law}.

\section{A Model}
\begin{figure}[t]
\figurenum{3}
\includegraphics[width=1.0\columnwidth]{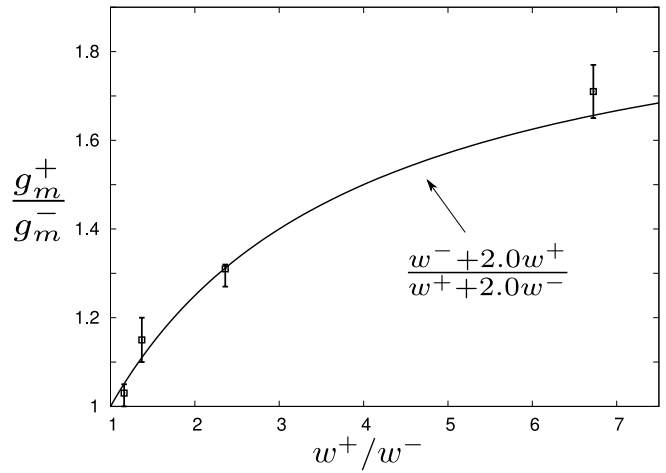}
\caption{The ratio of forward to backward diffusion as a function of imbalance.}
\label{law}
\end{figure}

Hydrodynamic turbulent diffusion forward and backward in time is known to be different by a factor of $a \approx 2$ \citep[see, e.g.,][]{berg2006},
which is due to fundamental non-reversibility of turbulence.
In our case the diffusion of magnetic field lines comes from the $w^+$ component, which propagates against mean field
and $w^-$ component which propagates along mean field. The diffusion of field lines along the field will be ``forward in time''
for $w^-$ and ``backward in time'' for $w^+$ and vice versa for the opposite direction. Assuming that the diffusion from $b^\pm$
is proportional to the amplitude $w^\pm$, we can write for the diffusion asymmetry:
\begin{equation}
\frac{g_m^+}{g_m^-}=\frac{w^-+a_mw^+}{a_mw^-+w^+},
\end{equation}

where $a_m$ is the time-asymmetry in MHD turbulence. Fig.~\ref{law} indicates that this model agrees with data reasonably well,
as long as $a_m=2.0\div 2.1$, which is compatible with hydrodynamic time-asymmetry.

\section{Relation to Goldreich-Sridhar anisotropy}
Eq.~2 looks similar to the critical balance of \citet{GS95}. Introducing anisotropy constant $C_A$, critical balance can
be written as $\ell_\|=C_A v_A \ell_\perp^{2/3} \epsilon^{-1/3}$ \citep{B12b}, where $\ell_\|$ is the distance paralle to the field, i.e.,
analogous to $x$ and $\ell_\perp$ is a perpendicular distance. This is indeed the same functional dependence as Eq.~2.
However, if the assume that these relations are identical, this would imply that $C_A=g_m^{-1/3}$. This is not satisfied, however.
The difference is that the measurement of the diverging magnetic fields is quasi-Lagrangian, while the measurement of
structure function that lead to anisotropy constant is Eulerian.
This difference becomes even more pronounced in the imbalanced case, where each of the $w^\pm$ components
has its own anisotropy. The analogy between Richardson-Alfven diffusion and critical balance 
would suggest that $g^+ (\ell_\|^+)^3=g^- (\ell_\|^-)^3$. We presented the ratio of parallel scales 
in the middle of the inertial range in Table~1. As we see, the expression above is not satisfied and  
the ratio of magnetic diffusions can not be explained by the anisotropy difference. So, despite the similar functional form,
Richardson-Alfv\'en diffusion has no direct relationship to Goldreich-Sridhar anisotropy.

\section{Discussion and Implications for Particle Transport}
Our measurement is the first demonstration of the $x^3$ superdiffusion of magnetic field lines in simulations of MHD turbulence.  
Superdiffusion of fast particles in the solar wind has been argued based on observational data \citep{Perri2009}.
Superdiffusion of field lines has been discussed in \citet{jokipii1973}.
The reconnection model of \citet{Lazarian1999} also uses perpendicular superdiffusion of field lines.
The superdiffusion of particles has been argued in \citet{narayan2001,lazarian2006,yan2008}, however, the asymmetric superdiffusion has not been anticipated before.
Our earlier measurement of perpendicular diffusion using MHD simulations \citep{BYL11} has been made in the large separation
limit and reproduced FLRW, which is symmetric. The measurements of cosmic ray propagation in artificial random fields,
such as \citet{Giacinti2012} can, in principle, reproduce superdiffusion, but since artificial fields lack the time-asymmetry of turbulent fields,
they can not reproduce asymmetric diffusion. Based on the similarity between Goldreich-Sridhar anisotropy and Richardson's diffusion \citet{narayan2001}
suggested that magnetic field lines separate within the Goldreich-Sridhar cone, however, according to the section above, this analogy is misleading,
especially in the imbalanced case. Time-asymmetry of turbulence, that we confirmed in this Letter, have consequences
for small-scale dynamo as well \citep{B12a}.

One of the consequences of asymmetric perpendicular diffusion is an induced streaming. Indeed, if we consider two close magnetic field tubes,
one of which is filled with isotropically distributed particles and another empty, the asymmetric diffusion into the empty tube will result
in an average streaming $\sim (1-g^+/g^-)$ of particle's velocity. In particular, for relativistic particles, such as cosmic rays,
this will result in streaming velocity of $2c(1-g^+/g^-)/\pi$, which could easily exceed the threshold for streaming instability, $v_A$
as long as imbalance amplitude $1-w^-/w^+$ exceeds $(3\pi/2) v_A/c$, which is around $10^{-4}$ in the WISM. Therefore, the induced streaming
will be counteracted by streaming instability \citep{kulsrud1969}. Above the threshold for turbulent damping \citep{farmer2004,BL08b}
streaming instability will be suppressed and, according to the estimates in the above paper, the weak large-scale streaming should reappear
at energies $10^{11}$ eV and strong streaming is expected above $3\cdot 10^{13}$ eV, although such energies are already
heavily influenced by pitch-angle scattering.
The net effect of the streaming instability from particles with energies below $3\cdot 10^{13}$ eV will be a flux of slab waves
which will increase the rate of pitch-angle scattering for these particles. The modeling of this effect will be subject of a future publication.

\ \ 

{\it Acknowledgments:}
AB was supported by Los Alamos Director's Fellowship.
\ \ 

\ \

\def\apj{{\rm ApJ}}           
\def\apjl{{\rm ApJ }}          
\def\apjs{{\rm ApJ }}          
\def\grl{{\rm GRL }}
\def\aap{{\rm A\&A } }
\def\mnras{{\rm MNRAS } }
\def\physrep{{\rm Phys. Rep. } }               
\def\prl{{\rm Phys. Rev. Lett.}} 
\def\pre{{\rm Phys. Rev. E}} 
\bibliography{all}

\end{document}